\documentclass[aps,pra,twocolumn,amsmath,amssymb,superscriptaddress,showpacs]{revtex4-1}

\usepackage{amsfonts}
\usepackage{amsmath}
\usepackage{amssymb}
\usepackage{amsthm}
\usepackage{fontenc}
\usepackage{graphicx}
\usepackage{xcolor}
\usepackage{textcomp}
\usepackage{epstopdf}
\usepackage{braket}
\usepackage{mathtools}
\usepackage{amsmath}
\usepackage{dcolumn}
\usepackage{multirow}
\usepackage{units}
\usepackage{gensymb}

\usepackage{soul}

\begin{document}

\title{Sliding-induced topological transitions in bilayer biphenylene}
\author{L. L. Lage}\email{lucaslage@id.uff.br}
\affiliation{Instituto de F\'isica, Universidade Federal Fluminense, Niter\'oi, Av. Litor\^{a}nea sn 24210-340, RJ-Brazil}

\author{S. Bravo}
\affiliation{Departamento de F\'isica, Universidad T\'ecnica Federico Santa Mar\'ia, Valpara\'iso, Chile}
\author{O. Arroyo-Gascón}
\affiliation{Nanotechnology Group, USAL-Nanolab, University of Salamanca, E-37008 Salamanca, Spain}
\author{Leonor Chico}
\affiliation{GISC, Departamento de F\'{\i}sica de Materiales, Facultad de Ciencias Físicas, Universidad Complutense de Madrid, E-28040 Madrid, Spain}
\author{A. Latgé}
\affiliation{Instituto de F\'isica, Universidade Federal Fluminense, Niter\'oi, Av. Litor\^{a}nea sn 24210-340, RJ-Brazil}

\date{\today}

\begin{abstract}

Sliding-induced topological transitions in biphenylene bilayers are investigated, considering various stacking configurations which are analyzed from a symmetry perspective and described in detail, highlighting the intricate patterns of type-II Dirac cone crossings.
Topological changes in the Fermi surface are assessed via the Euler characteristic,
linking each transition to its corresponding symmetry, which can be experimentally tested by conductance measurements. Moreover, the ability to tune these topological properties by sliding the layers provides a simpler and more effective way to observe such phenomena.     

\end{abstract}

\maketitle

\section{Introduction}\label{Introduction}

\begin{figure}[!h]
    \centering
\includegraphics[width=8.6cm]{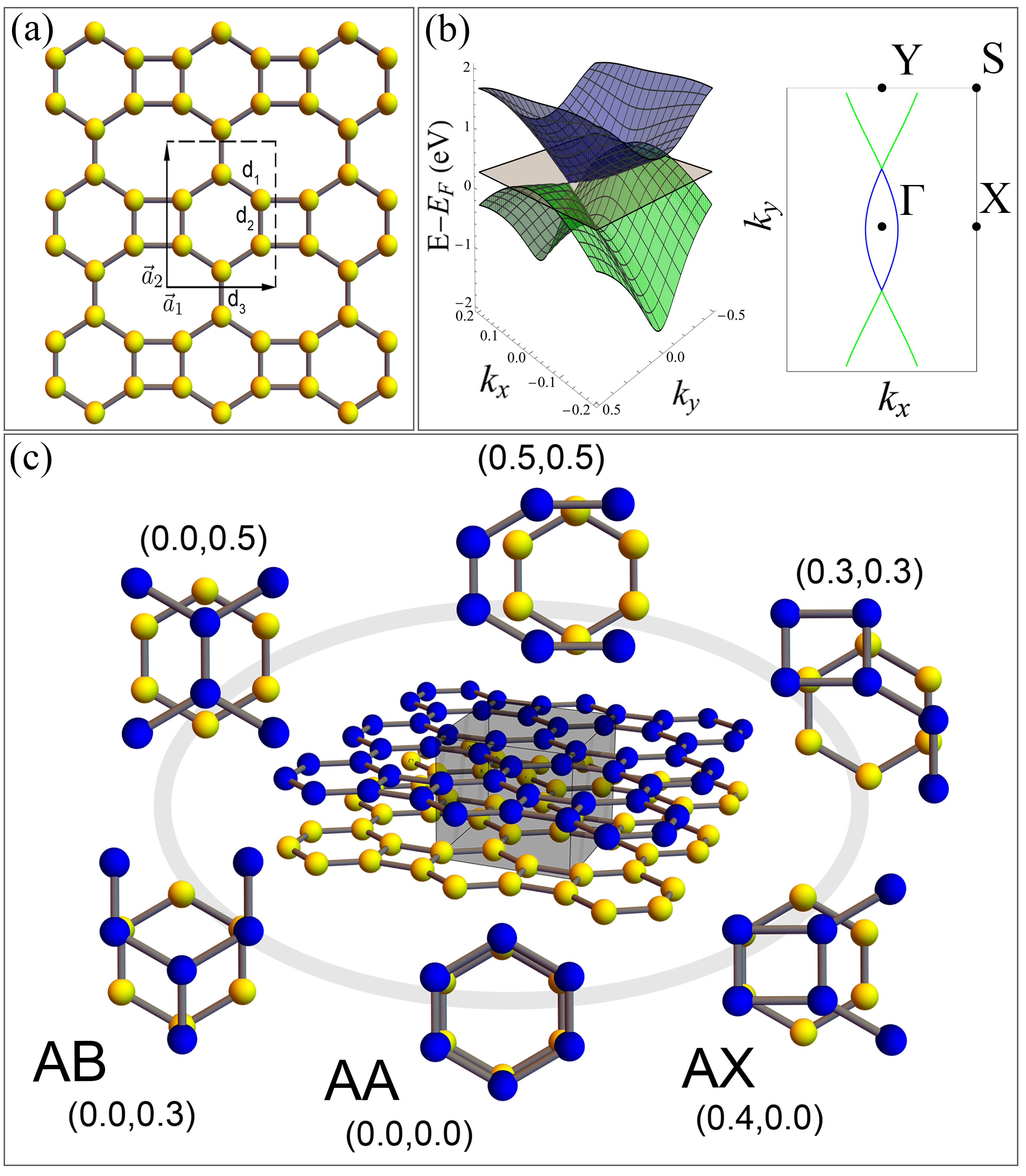}
\caption{(a) Biphenylene monolayer structure with bond lengths $d_1=1.41$\,Å, $d_2=1.45$\,Å, and $d_3=1.46$\,Å. (b) Type-II band structure of BPN monolayer with Dirac points at $E=0.26$\,eV; valence (green) and conduction (blue) bands are shown. (c) Bilayer stacking configurations labeled $(\delta_x, \delta_y)$.}\label{FIG1}  
\end{figure}

The biphenylene network (BPN) is a promising two-dimensional (2D) material that was predicted long ago \cite{Balaban1968} and has experienced a revitalization due to its unique structural and electronic properties \cite{hudspeth2010electronic}. 
Recently, it has gained significant attention mainly due to its experimental synthesis \cite{fan2021biphenylene}. As a derivative of biphenyl, biphenylene consists of two 
hexagonal carbon
rings connected directly by two carbon-carbon double bonds in a linear arrangement, resulting in a planar $sp^2$ structure that combines six-fold, eight-fold, and four-fold carbon rings [see Fig. \ref{FIG1}(a)]. Being a new 
carbon 
allotrope related to graphene but with a mixed geometry \cite{GIRAO2023611}, it possesses distinct electronic properties and symmetries. In particular, its band structure shows a type-II Dirac cone near the Fermi energy, with carrier velocities of the same sign and anisotropic transport properties \cite{Padilha2024}. These features have been proposed to be of interest for nanoelectronics \cite{Liu2021,Bafekry2022biphenylene}.
The electronic stability of one-dimensional biphenylene systems was also studied before its synthesis using first-principles calculations, including ribbons and tubes of different widths and morphologies \cite{hudspeth2010electronic}. 
Because of the complex geometry of biphenylene, novel stacking configurations may be achieved by piling up such carbon-thin sheets. In a recent work \cite{lage2024}, we proposed new symmetric bilayer stackings with different electronic properties.
Similarly to graphite, these bilayers are held together by van der Waals (vdW) forces.  In bilayer graphene, the two primary stacking arrangements of interest are AA and AB (Bernal), with different Dirac cone patterns, namely, four linear crossings or touching parabolas, respectively, at the Fermi level. By introducing a gate voltage between the layers, in the AB case a band gap can be achieved \cite{Castro2017}. Trigonal warping effects also modify the parabolic dispersion, with the appearance of sets of Dirac-like linear bands \cite{McCann2007}. 
For bilayer biphenylene, in contrast to the monolayer, the lower-energy type-II Dirac cone in the conduction band is split. In the AA case, this splitting gives rise to two cones, so that one of them is actually very close to the Fermi level. Each of these cones has two branches with velocities of the same sign and dissimilar magnitude, in contrast to the isotropic behavior of Dirac cones in graphene \cite{lage2024}.

Another way to tune the band structure properties of monolayer biphenylene is by means of strain: in addition to modifying the position of the Dirac cone, the Fermi surface characteristics can be altered, as recently shown \cite{Son2022}. In fact, the Fermi surface is a fundamental feature in understanding the electronic topology of a system. 
At some critical points, the band connectivity can change abruptly, as observed by Lifshitz long ago \cite{Lifshitz1960}.  
To detect such electronic topological transition, known as a Lifshitz transition \cite{Volovik2017}, the Fermi level can be adjusted to the exact place in the band structure where the topological shift occurs. However, in general, its experimental implementation involves large variations
in electron density by doping or by applying high pressure or magnetic fields
\cite{Galeski2022,VARLET201519,Aygar2024,PhysRevLett.112.146403,Chi2024}.
The rise of low-dimensional materials has allowed for sizeable modifications of such characteristics via strain. 
These changes can include the splitting or merging of Fermi surface pockets of both monolayer and bilayer systems \cite{PhysRevResearch.5.023120,VARLET201519,PhysRevB.84.155410} 
and have been experimentally observed in the electronic bands of graphene \cite{balseiro,Aygar2024}.
Topological transitions could dramatically change the physical properties of the materials, 
such as thermoelectric and transport responses \cite{doi:10.1021/acs.nanolett.0c03586}. Indeed, applied in-plane strain or displacements between layers in graphene bilayer systems may also change the topology of the Dirac cones \cite{PhysRevB.84.155410, VARLET201519,BHATTACHARYYA2016432}. 
 Also, the topological properties of graphene-based bilayer can be adjusted by sliding \cite{PhysRevB.84.155410} and twisting \cite{Andreireview,Fujimoto2021,PhysrevLett.128.026404} their layers. This points to the exploration of similar mechanisms for biphenylene-based systems.

Here, we propose novel bilayer configurations by continuously sliding one BPN layer over the other. We have found that such new stackings may have different band structures with respect to the split Dirac cones. 
 Additionally, we have identified several Lifshitz transitions induced by sliding, which are allowed without externally breaking the crystal symmetries or doping the 
 material.  
To investigate changes in the band structure and Fermi surface topology related to these transitions, we have performed first-principles Density Functional Theory (DFT) calculations, as detailed in the Supplementary Material (SM) \cite{supp}. Since spin-orbit effects
are negligible (of the order of meV) compared to typical hopping parameters (of the order of eV) in biphenylene systems, the effect is not included in the following computations, as previously discussed \cite{lage2024}. DFT calculations of monolayer biphenylene reveal a type-II Dirac cone with electron- and hole-like pockets \cite{soluyanov2015type} for a surface cutting the Dirac nodes, as shown in Fig. \ref{FIG1}(b). Moreover, Dirac nodes do not occur at high symmetry points, as also reported for bidimensional transition metal dichalcogenides (TMDs) such as WTe$_{2}$ and MoTe$_{2}$ \cite{PhysRevX.6.041069}. A tight-binding description fitting quite well with our DFT results is also discussed in Appendix A. 

Because the studied bilayer BPNs of different stacking profiles exhibit metallic properties, we must choose a topological invariant capable of capturing a topological transition, since it is not possible to directly observe a usual insulating topological transition from the corresponding band structures. We use the Euler
characteristic as a topological invariant to analyze the different stacking configurations. Our findings show distinct behaviors depending on their symmetry, which we call high-symmetry stacking (HSS) and low-symmetry stacking (LSS). In the case of the LSS set, the invariant does
not discriminate between different sliding configurations, since all yield a null invariant.



\section{System geometries and symmetries }

The geometry of monolayer biphenylene is shown in the left panel
of Fig. \ref{FIG1}(a). 
The periodicity of the system is described by a rectangular unit cell in
the $xy$ plane, centered on the hexagonal ring. Repetition of this pattern creates octagonal rings and four atom rings. The relaxed unit cell vectors
$\vec{a}_1=3.76 \ \hat{x}$ and $\vec{a}_2=4.52 \ \hat{y}$ are also shown
in the figure. 
Likewise, the remaining relaxed geometric parameters are the following: the interlayer distance is 3.39
\AA \ in the perpendicular direction; the three inner-layer distances are the same for all bilayers, $d_1=1.41$ \AA, $d_2=1.45$ \AA, and
\ $d_3=1.46$ \AA, which also coincide with those of the monolayer,
indicated in Fig. \ref{FIG1}(a). The corresponding Brillouin zone (BZ) is depicted in the right
panel [Fig. \ref{FIG1}(b)], showing the labels of the high-symmetry points, along with a constant energy cut which contains the type-II Dirac points.

When bilayers are formed, diverse stacking possibilities emerge.
This freedom is ultimately restricted by physical and chemical properties that
establish the most favorable configurations to be realized experimentally. 
In later sections we will comment on the stability and energetic
landscape that the different stackings comprise, but for the sake of symmetry 
analysis, we focus here on the relative spatial orientation of the atomic sites and their space group classification.

\begin{figure}[!h]
\centering
\includegraphics[width=7cm]{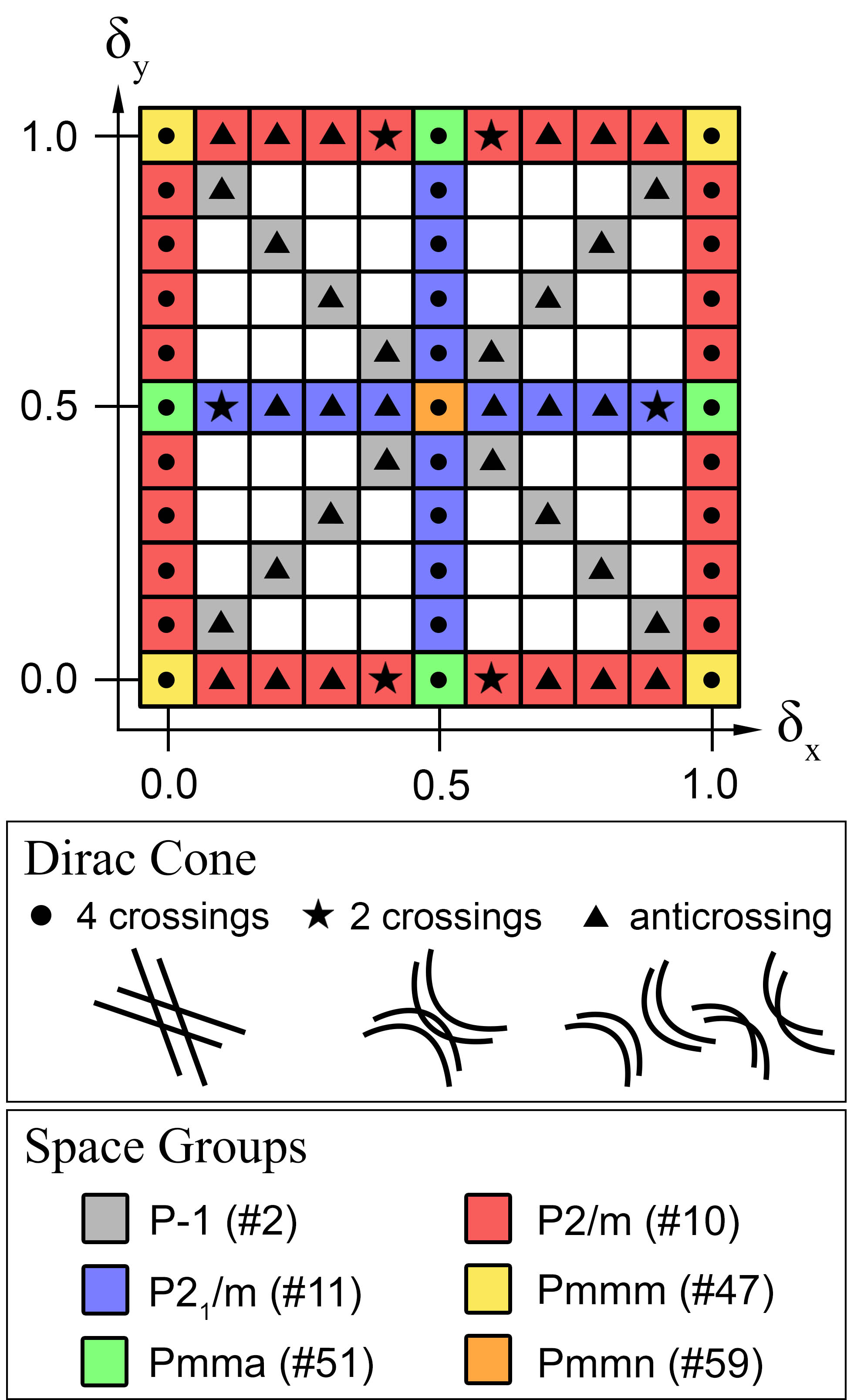}
\caption{Sliding diagram for the bilayer stackings: classification of the
different space groups and Dirac cone patterns with respect to the
displacements $\delta_x$ and $\delta_y$. Each square of the grid is
colored according to the space group of the specific bilayer; the color
code is shown in the bottom panel. Inside the square, a symbol indicates
the type of band crossing at the double Dirac cone closest to the Fermi
energy: circles (4 crossings), stars (2 crossings) and triangles (anticrossing),
as indicated in the panel below the diagram.}
\label{FIGdiagram}
\end{figure}

The coupling of two biphenylene monolayers can be performed following
different geometric configurations. First, we restrict the set of stackings in which rotational symmetries
play a role. Thus, we discard all stackings with the trivial space group
P1 (No. 1). We classify the remaining stackings into two main sets, 
introducing a useful notation.

Fig. \ref{FIG1}(c) presents some of the stackings studied in this work
(more configurations are detailed in the SM \cite{supp} \textcolor{black}{Fig. S1}). The stackings
are depicted by showing the hexagonal ring of the bottom layer
(yellow atoms) and the six atoms on the top layer (blue atoms) within
the unit cell. To generically describe the relative sliding, we fix the bottom
biphenylene layer and translate the upper one. We assume that the starting
position of the upper layer is the one where the in-plane atomic coordinates
coincide with the in-plane atomic coordinates of the bottom layer, i.e. direct or AA stacking. Rigid translations are described by a vector 
$\boldsymbol{\delta}=(\delta_x,\delta_y)$ in the $xy$ plane whose
Cartesian components quantify the amount of displacement in fractions of
the respective lattice vectors, so that $\delta_x$ and $\delta_y$ can take
values from 0 to 1. It is important to note that the sliding can be described in this form because the same rectangular unit cell (with 12 atoms) can be
maintained throughout the process.

The three stackings shown in Fig. \ref{FIG1}(c), denoted AA
\cite{Chowdhury2022}, AX, and AB, were reported by us in a previous work
\cite{lage2024}. In this work we explore the continuum of geometries obtained by smoothly sliding one biphenylene
layer on top of the other. As an example of the novel systems studied here, the three cases depicted in the upper part of the panel
are analyzed here for the first time. All the bilayers are identified in Fig. \ref{FIG1} (c)
with their corresponding coordinates $(\delta_x,\delta_y)$. Using this vector notation, two sets of stackings can be described. Starting
the sliding process at $\boldsymbol{\delta}=(0,0)$, we find three possible
routes with non-trivial rotational symmetries. We can follow the lines $(\mu,0)$
and $(0,\mu)$, or the diagonal line $(\mu,\mu)$, with $\mu$ ranging
from 0 to 1. For simplicity, we allow $\mu$ to take values only from $0$
to $0.5$, since the other quadrants of the $(\delta_x,\delta_y)$ plane are
obtained by symmetry considerations.

First, we focus on four possible stacking configurations with the maximal number
of symmetries that biphenylene bilayers can have, namely eight.
This  high-symmetry stacking set comprises the
$(0.0,0.0)$, $(0.5,0.0)$, $(0.0,0.5)$, and $(0.5,0.5)$ geometries.
All rotational symmetries of the HSS set
can be connected to the point group mmm ($D_{2h}$), the same as in the HSS set. 
that generates
the monolayer space group. The difference among the space groups
of the HSS set is that for the stackings $(0.5,0.0)$, $(0.0,0.5)$ and
$(0.5,0.5)$, some of the rotational symmetries become nonsymmorphic.
In particular, for $(0.5,0.0)$ and $(0.0,0.5)$ stackings, four of the eight
symmetries become nonsymmorphic with a fractional translation vector
equal to the corresponding sliding vector. In the
case of the $(0.5,0.5)$ stacking, four symmetries also become nonsymmorphic, 
but now the fractional translation vector has both nonzero components. 
In summary, with respect to the HSS set, the $(0.0,0.0)$ stacking is characterized by the space group
(SG) Pmmm (No. 47), the $(0.5,0.0)$, $(0.0,0.5)$ stackings by SG Pmma (No. 51),
and finally the $(0.5,0.5)$ stacking by SG Pmmn (No. 59).
The low symmetry stacking (LSS) set is composed of configurations
$(\mu,0)$, $(0,\mu)$ and $(\mu,\mu)$ [with $\mu \in
(0.0,0.5)$], as well as two additional lines with rotational
symmetries which arise by fixing one of the components of the $\boldsymbol{\delta}$
vector to $0.5$ and varying the other component, i.e. $(\mu,0.5)$, $(0.5,\mu)$.
For this set, the classification is also straightforward. When we move
along the $(\mu,0)$, $(0,\mu)$ lines we break four symmetries; 
only spatial inversion, a rotation, and a mirror plane, both with respect to the
axis in which the sliding occurs, are preserved. This yields SG P2/m
(No. 10) for both sliding lines. 
In the case of the $(\mu,0.5)$, $(0.5,\mu)$ lines, the number of rotational symmetries 
is the same as in the previous 
cases, but with
nonsymmorphic rotation and 
mirror symmetries. This yields SG P$2_1$/m
(No. 11) for both lines. Finally, diagonal sliding breaks all symmetries
with the exception of spatial inversion, resulting in SG P$\Bar{1}$ for
the $(\mu,\mu)$ line.
\begin{figure*}[t!]
\centering   
\includegraphics[width=\textwidth]{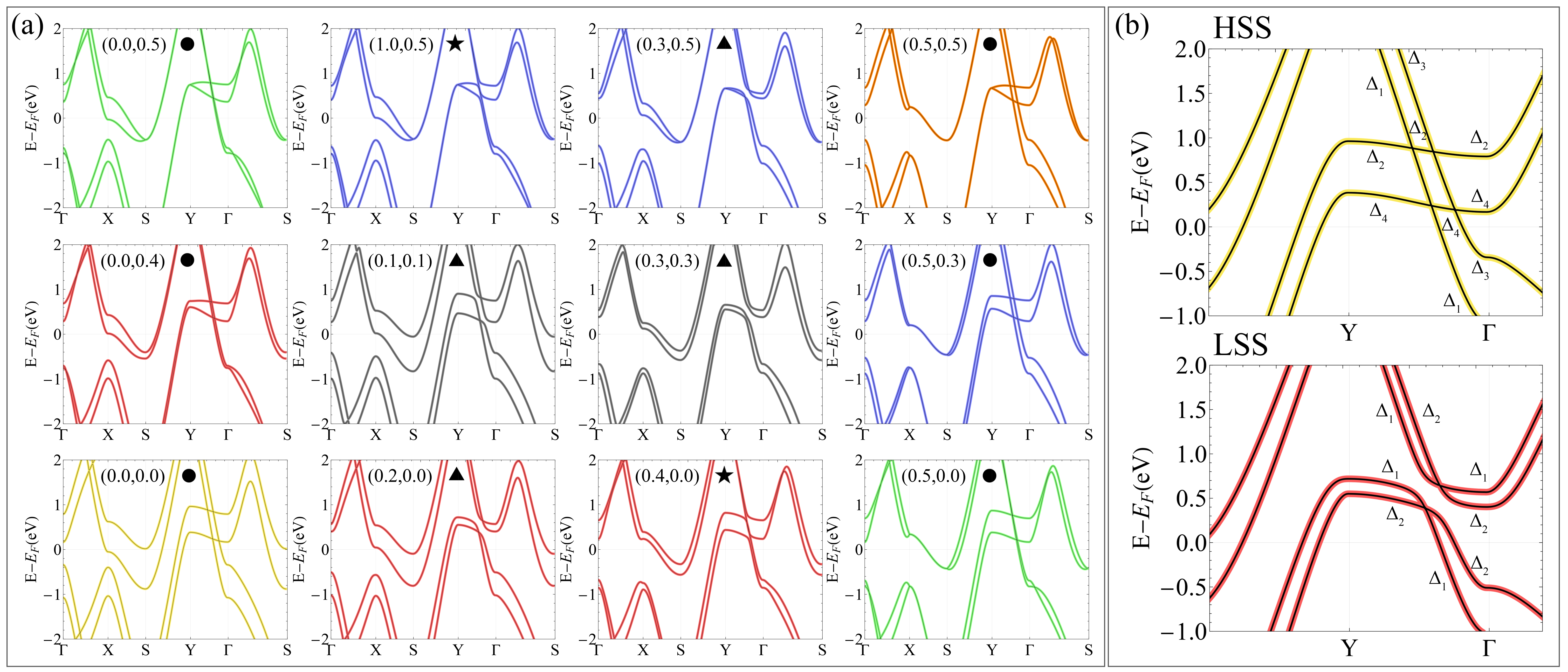}
\caption{(a)
Band structures for the biphenylene bilayers corresponding to the displacements $(\delta_{x},\delta_{y})$. Symbols and colors refer to the
diagram of Fig. \ref{FIGdiagram}. 
(b) Electronic bands along the $Y$-$\Gamma$ line exhibiting the four
crossings for HSS $(0.0,0.0)$ and LSS $(0.2,0.0)$, labeled with the corresponding irreps $\Delta_i$.  }
\label{FIGbands}
\end{figure*}
The relation between the symmetry and the geometry of the different
sliding configurations with respect to the layer displacements
$\delta_x$ and $\delta_y$ is graphically summarized in a
chessboard-like diagram in Fig. \ref{FIGdiagram}. 
Each bilayer configuration belongs to a space group that may
change with displacement. The symmetry of the bilayers is
visually represented by the color of the squares composing the
diagram. 
We show in the following that the diversity of symmetries presented by the BPN bilayers
is closely related to their different band structures. Moreover, the band crossing and
anticrossings at the type-II Dirac cones along the $Y-\Gamma$
path present three possibilities: four crossing bands, two crossing bands
and two types of anticrossing bands. They are represented by circles, stars
and triangles, respectively, inside the diagram boxes of Fig. \ref{FIGdiagram}.

\section{First-principles structural and electronic properties}

We perform first-principles calculations for all HSS and for a subset
of LSS which correspond to values of $\mu = n/10$ for $n \in
\{1,2,3,4\}$. For the bilayer structures studied in this work, the
sliding process does not produce substantial structural modifications
such as unit cell volume variations or interlayer distance fluctuations.

Fig. \ref{FIGbands}(a) gathers the band structures corresponding to
representative cases from the sliding diagram in Fig. \ref{FIGdiagram}.
Notice that the complete symmetry information for each sliding
configuration is included in the first quadrant region of the
diagram, which includes the displacements $0 \leq \ \delta_{x,y} \
\leq 0.5$. Thus, the bands presented in Fig. \ref{FIGbands} are colored
according to the space group code for this region of the diagram.
Concerning the electronic structure and its dependence on sliding, we start
by analyzing the HSS set. The energy bands near the Fermi level for
these four stackings are presented in the four corner panels of Fig.
\ref{FIGbands} (a).

The bands for the direct stacking case AA [$(0.0,0.0)$; SG No. 47] are
presented at the bottom left corner of the figure (yellow curves).
The bands are completely split and cross at four points. 
The two main features discussed are the Dirac nodes and the characteristics
associated with the nonsymmorphic symmetries. 
Considering the latter, the most notable feature due to the joint action
of a nonsymmorphic operation and time-reversal symmetry is the formation
of nodal lines along the boundaries of the BZ; the so-called stick-together
bands \cite{dresselhaus2008group}. The type of nodal line is related to
the type of stacking. Thus, we can see that the case $(0.5,0.0)$ has a nodal
line along the $X-S$ high-symmetry path, and the case $(0.0,0.5)$ has a nodal
line along $Y-S$.
Both cases correspond to the space group No. 51, and their band structures are
shown in green at top-right and botton-left panels of Fig. \ref{FIGbands} (a).
As expected, the $(0.5,0.5)$ bilayer presents a nodal line along the entire
boundary of the Brillouin zone. Therefore, the HSS set allows us to study the entire range of boundary nodal lines.

The presence of type-II Dirac cones can be directly observed in the
band structures. All HSS examples host four cones along the $\Gamma-Y$
line in reciprocal space. The cones are produced by the crossing of two pairs of bands with large magnitude of the velocity with another pair with much smaller velocities. 
Each pair stems from the bilayer splitting. The location of the crossings is modified by varying the stacking: the $(0.5,0.0)$ and $(0.0,0.5)$ cases present a very close pair
of Dirac cones while the other two configurations have a higher separation between the cones. In terms of reciprocal space symmetry, the superposition of the bands
actually entails crossovers by analyzing the irreducible representations
of the bands along the $\Gamma-Y$ line. From a DFT-based group theory analysis 
\cite{Iraola2022,Aroyo2006,Aroyo2006-2}, the symmetry along this line is
described by the point group isomorphic to mm2 (C$_{2v}$) with
four elements. This is true for all the HSSs, and implies that there are four irreducible representations describing the symmetry of the bands in this
region. 

We study the symmetry of the bands along $\Gamma-Y$ (the so-called $\Delta$
line) at different locations, and present the results in Fig. \ref{FIGbands} (b),
for the representative HSS $(0.0,0.0)$ and LSS $(0.2,0.0)$ geometries.
For the HSS case, the irreducible representations of the bands that meet at each of the nodes
are different, 
and thus cannot mix, indicating that they
can only cross. This implies that these crossings are strict Dirac points.
Now we turn to the LSS. We focus on the subset of lines
$(\mu,0)$ and $(0,\mu)$ (red bands in Fig. \ref{FIGbands}), since the
lines $(\mu,0.5)$ and $(0.5,\mu)$ (blue bands) behave
similarly.
Additional bands for
these two lines are also presented in 
Fig. \textcolor{black}{S2-S3} of the SM \cite{supp},
where the first row is for specific values on the $(\mu,0)$ line and the
second row represents the cases for the $(0,\mu)$ line.
In general terms, moving away from the HSS set, the momentum-space symmetries are reduced along the $\Delta$ line, decreasing the number of available irreps to only two. Therefore, there are fewer possibilities to form four
simultaneous Dirac nodes. Namely, four nodes are possible only if the 
pairs of crossing bands transform under 
different irreps. If each pair of bands with different velocities has a mixed irrep character, i.e. one irrep coincides and the other is different, then there can be at most two crossings along $\Delta$.
All scenarios (four, two, and zero nodes) are schematically represented in the diagram presented in Fig. \ref{FIGdiagram} and examples of the band structure are shown in Fig. \ref{FIGbands}.  
It can be observed from the diagram that the cases with four crossings occur along the $(0,\mu)$ line. 
One of these, the $(0.0,0.0)$ stacking, is shown in the upper panel of Fig. \ref{FIGbands}(b). 
All the irreps of the bands at the double Dirac cone are different, so they actually cross. 
This irrep configuration is respected along the entire $(0,\mu)$ line, which explains the robustness of the four Dirac crossings along it.

A different scenario appears along the $(\mu,0)$ line, 
where, due to the order of the irreducible representations, the bands at first do
not cross and from approximately $\mu = 0.3$ two crossings appear
(see Fig. \ref{FIGbands}(a) and Fig. \textcolor{black}{S2} in SM \cite{supp}). 
The anticrossing behavior can be understood by resorting to the lower panel of Fig. \ref{FIGbands}(b), which corresponds to the $(0.2,0.0)$ stacking. The pair of bands with higher velocity have irreps $\Delta_1$ and $\Delta_2$, the same as the pair of bands with lower velocity. Due to the ordering of the bands, they are forced to anticross, so the Dirac cone is gapped. If the irrep order of one pair of bands at the Dirac point were swapped, crossing would be allowed. 

These crossings remain robust until the HSS $(0.5,0.0)$ is reached.
This exhausts all possible arrangements of crossings in these bilayers. The behavior
on this line makes it clear that the appearance of Dirac nodes is not
enforced by symmetry, but only allowed by the underlying group structure.
In fact, we have shown that the node formation is highly dependent on the
irrep ordering, which is ultimately linked to the energetic properties of
stackings.
For the case of the diagonal line $(\mu,\mu)$, the symmetry analysis shows
that along the $\Delta$ line the symmetry is reduced to the trivial
group, and thereby only one type of irrep is possible. 
Examples of bands corresponding to $(\mu,\mu)$ stacking are presented
in the central panels (black lines) of Fig. \ref{FIGbands} and detailed
in Fig. \textcolor{black}{S4} in the SM \cite{supp}.
Therefore, no crossings are allowed along this line, hindering
the Dirac node formation.

\begin{table*}[t!]
\centering
\caption{Summary of Euler characteristics for the Fermi surface topologies presented in Fig. \ref{Fermi_surface_HSS}(a)-(d).}
\label{tab:euler_summary}
\begin{tabular}{|c|c|c|c|c|}
\hline
\textbf{Case} & \textbf{Coordinates} & \textbf{Blue curves (electron-like)} & \textbf{Red curves (hole-like)} & \textbf{Total $\chi$} \\
\hline
(a) & $(0.0, 0.0)$ & Open loop ($\chi = 0$) & Two loops ($\chi = -1$ each) & $-2$ \\
\hline
(b) & $(0.5, 0.0)$ & Two loops ($\chi = +1$ each) & Two loops ($\chi = -1$ each) & $-1$ [or $0$ along $(\mu, 0)$] \\
\hline
(c) & $(0.0, 0.5)$ & Open loop ($\chi = 0$) & One connected loop ($\chi = -1$) & $-1$ [or $0$ along $(0, \mu)$] \\
\hline
(d) & $(0.5, 0.5)$ & One connected loop ($\chi = +1$) & One connected loop ($\chi = -1$) & $0$ [also 0 along $(\mu, \mu)$]\\
\hline
\end{tabular}
\end{table*}
\section{Fermi sea topology}

The metallic character of all stackings motivates the study of the
connection between different configurations. We focus on the topology
of the complete Fermi sea and compute the Euler characteristic as a topological
invariant. We aim to present a robust response that can have different values
for the stackings.
The Euler characteristic is an invariant that describes the global
topology of a space \cite{nakahara2018geometry}. Considering in general
the Fermi sea as a disconnected space, which is embedded in the
BZ torus, there are several ways to compute this invariant. We follow Ref. 
\cite{PRB.107.245422_FStop} and use the definition  valid for 2D systems,
\begin{equation}
    \chi = \sum_{k} (2 - 2g_k - b_k)\,\,,
\label{EulerChar}
\end{equation}
\noindent 
where $k$ labels each different disconnected components of the Fermi sea,
$g_k$ represents the corresponding genus, and $b_k$ denotes the number of
its boundaries \cite{PRB.107.245422_FStop}. 
In general, the Fermi sea topology includes information from all
the bands crossing the Fermi level; 
these regions can be open or closed loops
in the torus that constitutes the BZ. Closed loops give rise to electron or hole pockets 
around certain regions of the BZ. 
Only closed loops contribute to the Euler characteristic magnitude;
neither open loops nor fully filled or unoccupied bands play a role in the Fermi sea topology. 

\begin{figure}[!h]
\centering   
\includegraphics[width=9
cm]{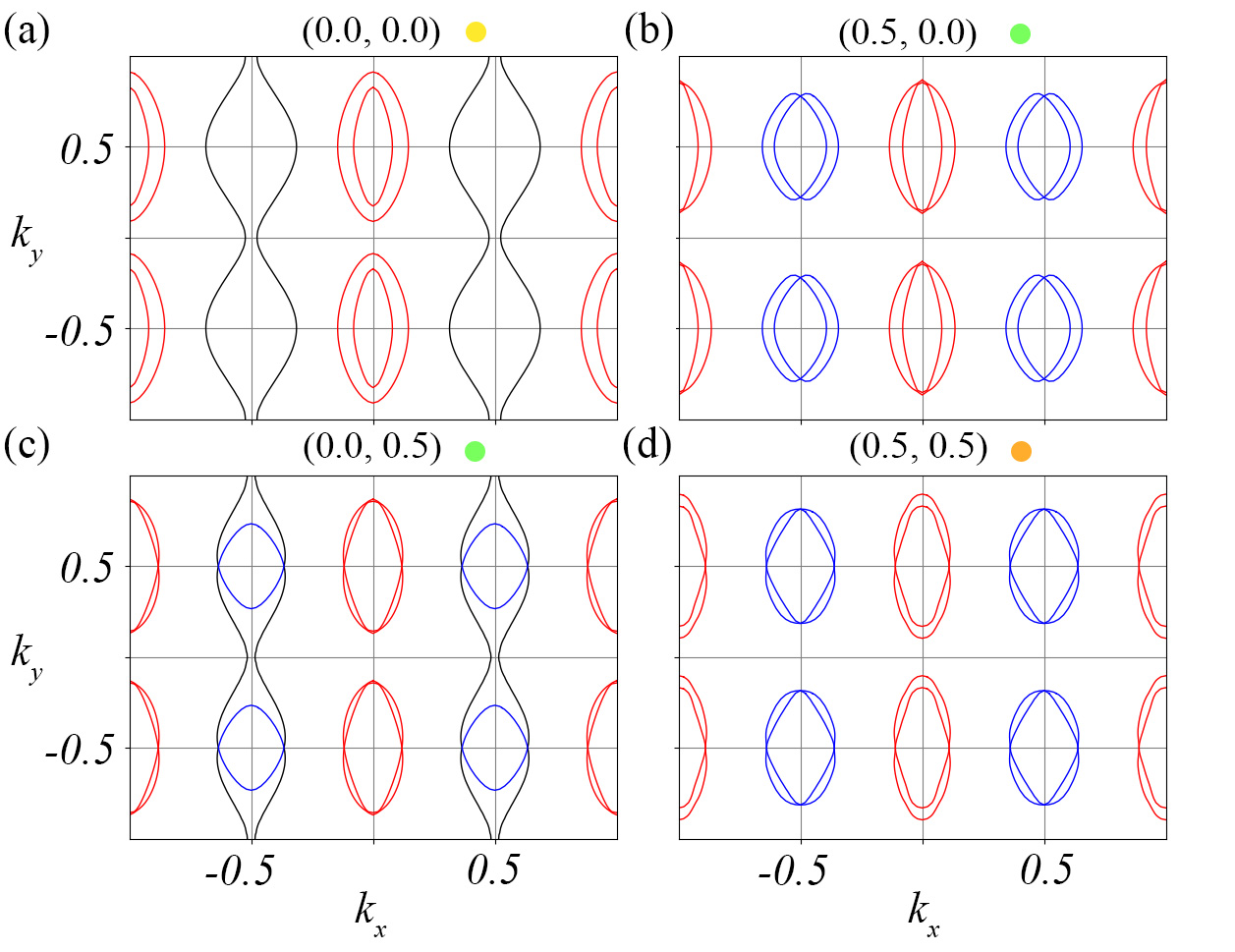}
\caption{Fermi surface (loops) for the HSS set. Blue and red solid lines
represent electron-like and hole-like surfaces, respectively. Each stacking is accompanied by the corresponding
group symmetry and Dirac cone type as colored circles.}
\label{Fermi_surface_HSS}
\end{figure}

In the following, we mainly restrict the analysis to the HSS set, although we
briefly comment on the results for the LSS set, which yields a null topological
invariant. We start with the $(0.0,0.0)$ stacking as shown in Fig.
\ref{Fermi_surface_HSS}(a). An open Fermi loop appears with an electron
character (blue solid line) and two closed loops with hole character
(red solid lines). Since the open loop does not contribute to $\chi$ and each
hole-like closed loop contributes with $-1$, the total result is $\chi = -2$.
The zero value of the open loop (labeled by $k = 1$) is obtained by 
verifying that $g_1 = 0$ and the boundary contribution $b_1 = 2$. 
Next, each
hole-like loop constitutes a separate input for $\chi$, named as $k=2$ and
$k=3$ components. Their genus values are $g_2 = g_3 = 1$ and $b_2 = b_3 = 1$.
Therefore, $k=2$ and $k=3$ contribute to $\chi$ with $-1$ each,
leading to a total value of $\chi = -2$.

Now we examine the $(0.5,0.0)$ stacking, as presented in 
Fig. \ref{Fermi_surface_HSS}(b). We note that the aforementioned open loop transitions to a closed loop, and a second closed
electron-like loop appears and merges with the initial loops. These two
loops are then connected due to the nodal line formation along the
$X-S$ line. In contrast, the hole-like part is still composed of two
disconnected loops. 
The combination of the information from the two types of loops produces a
Euler characteristic of $\chi = -1$. This is because
the electron-like connected loop contributes with $+1$, and the
disconnected hole-like loops contribute with $-1$ each. 
%
%

It is worth mentioning that the transitions described above, i.e. moving from
open to closed loop and the appearance of the new electron-like loop, 
occur along the $(\mu,0.0)$ line. Thus, there is a change in topology in this
region, which is characterized by a change in $\chi$.
Throughout this region, the hole-like loops are not connected, and thus they
contribute separately. The same holds for the electron-like loops. 
Therefore, the total invariant yields zero ($\chi=0$), which is
preserved along the entire line, since no additional topological
transitions occur. Representative Fermi surfaces graphs for this line and other cases are
detailed in the SM  \cite{supp} in Figs. \textcolor{black}{S[5-7]}. Hence, on sliding from the $(0.0,0.0)$ 
to the $(0.5,0.0)$ stacking along $(\mu,0.0)$, the sequence of the 
$\chi$ values is $\{-2,0,-1\}$.

The Fermi surface for the $(0.0,0.5)$ HSS example is depicted in Fig.
\ref{Fermi_surface_HSS}(c). On one hand, the electron-like contours merge in one
component, which is an open loop, so it does not contribute to the
invariant. On the other hand, the hole-like loops exhibit one connected component
due to the emergence of a nodal line, giving $\chi=-1$. Therefore, 
the final Euler characteristic of the complete Fermi surface is $\chi = -1$. As in the previous line $(\mu,0)$,
the $(0,\mu)$ line also gives a null value for $\chi$. Here again, 
the contributions of the electron-like and hole-like components compensate. The results
illustrating the behavior along this line are presented in the SM \cite{supp} at \textcolor{black}{Fig. S[4,5]}.
This way, sliding from the $(0.0,0.0)$  to the $(0.0,0.5)$ stacking,
along $(0,\mu)$, produces the same sequence of values for $\chi$, $\{-2,0,-1\}$.

Finally, for the $(0.5,0.5)$ stacking shown in Fig. \ref{Fermi_surface_HSS}(d), merges occur in both electron and hole-like loops. Thus, there is one connected component in each case; their contributions are canceled and
the Euler characteristic is zero. The topology and Euler characteristic for
the $(0.5,\mu)$, $(\mu,0.5)$ can be obtained by analyzing Figs. \textcolor{black}{S6} in the SM
\cite{supp}.

An important point regarding the application of the invariant to distinguish
different stacking phases is related to the change of $\chi$ with
respect to sliding. 
The Euler parameter enables us to differentiate between the HSSs $(0.0,0.0)$, $(0.5,0.5)$ and the pair $(0.0,0.5)$ and $(0.5,0.0)$, although it does not distinguish between the last two stackings (both have $\chi=-1$).
Sliding towards $(0.5,0.5)$ 
$\chi$ cannot identify a topological transition related to the change from the LSS lines to the HSS, since its value is always zero. This indicates that the
intrinsic connectivity of the loops at the Fermi level (considering, for example, nodal structures) is not completely represented in the Fermi sea topology described
by $\chi$, although it is a helpful complementary information.

In summary, the Euler characteristic can be used to distinguish some of the
HSS stackings with respect to the LSS. It is worth mentioning that
it has been correlated with transport responses related to nonlocal conductance at a planar Josephson junction \cite{PRL.130.096301_FStop}, and with conductance
assuming quantized values directly related to the Euler characteristic 
\cite{PRL_128.076801_QNLCond}. This establishes a 
direct correlation between an experimentally measurable quantity and a subset of our presented HSS phases.

\section{Conclusions}

In this study, we investigate sliding-induced topological transitions in bilayer biphenylene by exploring different stacking configurations. Through first-principles calculations and symmetry analysis, we demonstrate that the electronic properties of BPN bilayers can be significantly altered by sliding one layer over the other, leading to changes in the topology of the Fermi surface and the emergence of type-II Dirac cones.
Our findings reveal that the symmetry of the stacking configurations plays a crucial role in determining the electronic band structure and the features of the Dirac cone crossings. High-symmetry stackings exhibit distinct patterns of Dirac cone crossings, while low-symmetry stackings show reduced symmetry and fewer crossings. We employ the Euler characteristic as a topological invariant  to effectively capture the topological transitions in the Fermi surface, distinguishing between different stacking configurations. In particular, the Euler characteristic changes as the layers slide, providing a robust indicator of topological transitions in several cases.
We also observe that the formation of Dirac nodes is highly dependent on the irreducible representations of the bands, which are linked to the underlying symmetry and energetic properties of the stackings. This dependence highlights the intricate relationship between symmetry, band topology, and electronic properties in BPN bilayers.
The ability to tune the topological properties of BPN bilayers via sliding offers a promising route for controlling the electronic behavior in nanoscale devices. This tunability, combined with the unique electronic properties of BPN, highlights it as a potential candidate for applications in nanoelectronics and quantum computing.

\section*{Acknowledgments}
The authors would like to thank the INCT de Nanomateriais de Carbono for providing support on the computational infrastructure. LLL thanks the CNPq scholarship. AL thanks the FAPERJ under grant E-26/200.569/2023. LC acknowledges financial support from the Agencia Estatal de Investigación of Spain under grant PID2022-136285NB-C31, and from grant (MAD2D-CM)– (UCM5), Recovery, Transformation and Resilience Plan, funded by the European Union - NextGenerationEU. OAG acknowledges funding from European Union NextGenerationEU/PRTR project Consolidación Investigadora CNS2022-136025.

\appendix 
\setcounter{figure}{0}
\renewcommand{\thefigure}{A\arabic{figure}}

\section{Tight-binding formulation}

\begin{figure*}
    \centering
    \includegraphics[width=0.75\textwidth]{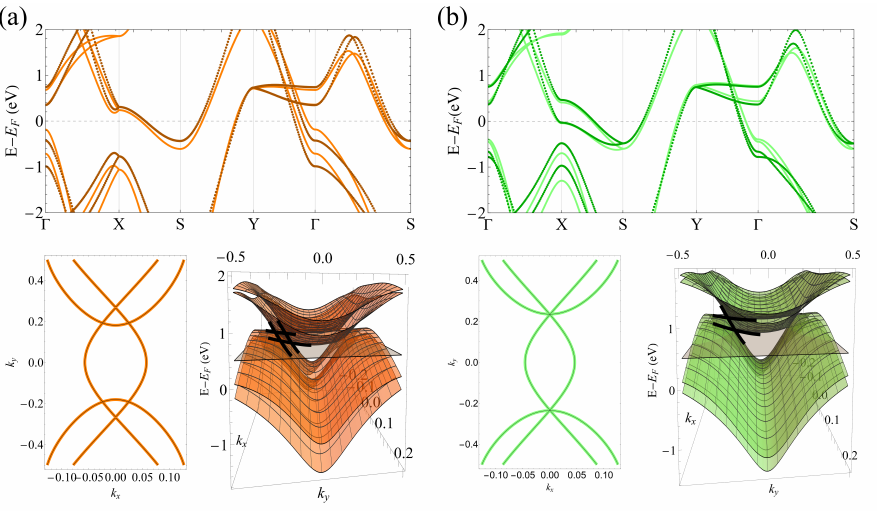}
        \includegraphics[width=0.75\textwidth]{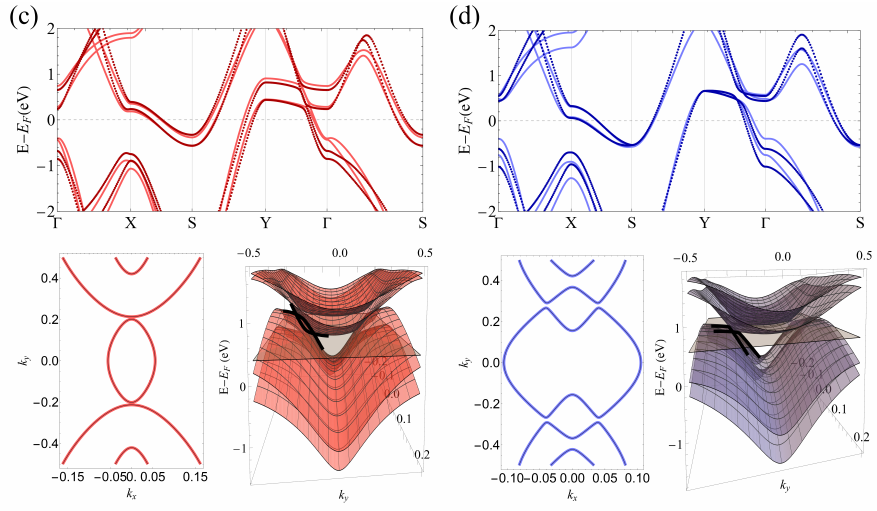}
    \caption{DFT (darker dots) and TB (solid lines) electronic bands  for different bilayer stackings, TB energy contours evaluated at the first cone crossing and the respective surface cone plot for the stackings (a-b) HSS $(0.5,0.5)$ and $(0.0,0.5)$, and (c-d) LSS $(0.4,0.0)$ and $(0.3,0.5)$, respectively.}
    \label{figbands}
\end{figure*}

A tight-binding parametrization is presented and compared with the DFT results. Our main goal is to examine the electrical characteristics of the bilayer structures for energies close to the Fermi level with low computational cost.  Since the biphenylene structure is composed of simple $sp^2$ carbon bonds \cite{fan2021biphenylene}, a single $p_z$ orbital tight-binding (TB) Hamiltonian is adopted, to describe the different stackings, given by

\begin{gather}
H=\sum_{i,a}\varepsilon_{i}^{a}c_{i}^{\dagger a}c_{i}^{a}+\sum_{\substack { i,j\\ {a} }}
t_{ij}^{a}c_{i}^{\dagger a}c_{j}^{a}+\sum_{\substack { i,j \\ {a \neq b} }} t_{ij}^{\perp a b}c_{i}^{\dagger a}c_{j}^{b}+h.c.,\label{Hamito}
\end{gather}

\noindent where $\varepsilon_i^{a}$ is the onsite energy of the atom located at the $i$  site in the layer $a$, and $c^{a\dagger}_{i}$ ($c^{a}_{i}$) creates (annihilates) an electron at site $i$ and layer $a$. The second term describes the intralayer couplings,
$t^{a}_{ij}$ being the corresponding hopping energies within the layer $a$. Clearly, for monolayers $a=1$, and the third summation is omitted. For bilayers, $a(b)=1,2$, and interlayer interactions, $t_{ij}^{\perp a b}$, depend on the stacking configuration between the top and bottom biphenylene layers. A suitable hopping parametrization is given by an intralayer hopping energy described by a decaying exponential function \cite{ lage2024}, 
\begin{equation}
    t^{a}_{ij}=t_1e^{-\beta\big(\frac{r_{ij}}{d_1}-1\big)} \ ,
\end{equation}
with $r_{ij}$ being the distance between $i,j$ lattice sites, $t_1$ the hopping related to the  
first 
nearest-neighbor distance $d_1$, and $\beta$ a fitting parameter that controls the range of the interaction. 
As the ratio $r_{ij}/d_1$ is always larger than one beyond the first nearest neighbors, small $\beta$ values allow one to increase the number of neighbors with non-negligible hoppings in the description.

For interlayer connection, we have also considered a decaying exponential function for the hopping energies given by $t_{ij}^{\perp a b}=t_0 e^{-\alpha \big(\frac{r_{ij}}{d_{\perp}}-1\big)}$
with $d_{\perp}=3.56$ \AA \ the smallest interlayer distance and $t_0$ the direct stacking hopping value when two carbon atoms are exactly one above the other.  The parameter $\alpha$ modulates the hopping strength between layers with increasing distance. The adopted TB parametrization is chosen by comparing the results with DFT calculations: $t_1 = -3.3$ eV, $t_0 = -0.33$ eV, $\beta = 2.2$, $\alpha = 1.47$. The onsite energy values are grouped into two sets: the four sites at the lateral corners of the hexagon and the two at the top and bottom vertices, given by $\epsilon_1=-1.8$ eV and $\epsilon=-2.2$ eV, respectively.

DFT and TB results of different stacking configurations are shown in 
Fig.~\ref{figbands}(a)-(d). Panels (a) and (b) correspond to two HSS, while panels (c) and (d) present two LSS cases. Each panel shows the band structures calculated with DFT (dark dots) and tight-binding (continuous lines) along the high-symmetry path, with different behaviors at the type-II Dirac cones. 
The lower left figure of each panel presents an energy TB contour plot
calculated at the energy of the first Dirac crossing (or anticrossing), next to the $\Gamma$ point.
The lower right figure of the panel shows a 2D rendering of the bands along with the constant energy plane
(in shaded gray) corresponding to the energy of the contour plot displayed in the bottom left part. 

Different crossing and anticrossing features are observed. The contour plots corresponding to the HSS shown in Figs.~\ref{figbands}(a) and (b), which belong to the $(0.5,0.5)$ and $(0.0,0.5)$ stackings respectively, have two crossings and one tiny anticrossing at each type-II Dirac cone. Figs.~\ref{figbands}(c) and (d), showing the results for LSS $(0.4,0.0)$ and $(0.3,0.5)$, respectively, reveal complete anticrossings between the contour lines.

\bibliography{refs}

\end{document}


\title{Supplemental Material: Sliding-induced topological transitions in bilayer biphenylene}
\author{L. L. Lage}
\affiliation{Instituto de F\'isica, Universidade Federal Fluminense, Niter\'oi, Av. Litor\^{a}nea sn 24210-340, RJ-Brazil}
\author{S. Bravo}
\affiliation{Departamento de F\'isica, Universidad T\'ecnica Federico Santa Mar\'ia, Valpara\'iso, Chile}
\author{O. Arroyo-Gascón}
\affiliation{Nanotechnology Group, USAL-Nanolab, University of Salamanca, E-37008 Salamanca, Spain}
\author{Leonor Chico}
\affiliation{GISC, Departamento de F\'{\i}sica de Materiales, Facultad de Ciencias Físicas, Universidad Complutense de Madrid, E-28040 Madrid, Spain}
\author{A. Latgé}
\affiliation{Instituto de F\'isica, Universidade Federal Fluminense, Niter\'oi, Av. Litor\^{a}nea sn 24210-340, RJ-Brazil}

\date{\today}

\maketitle

\subsection{Computational details}

 Density functional theory calculations were carried out using the Quantum ESPRESSO software
\cite{giannozzi_quantum_2009,giannozzi_advanced_2017} within the generalized
gradient approximation (GGA) and the Perdew-Burke-Ernzerhof (PBE) exchange correlation
functional \cite{Perdew1996}. 
In a first stage, a relaxation process was performed starting with the stacking of $(0.0,0.0)$. 
To achieve the other stackings, rigid displacements were used for the in-plane 
coordinates, simulating a clamp to obtain the desired sliding step. 
The out-of-plane coordinate was allowed to relax in order to minimize with a variable cell configuration. The force tolerance along the vertical
direction was set to 0.01 eV/\AA. 
After the relaxation stage, the electronic structure was obtained for each stacking, 
using an energy cutoff of 100 Ry, along with a tolerance of 10$^{-9}$ eV and a
norm-conserving pseudo-potential \cite{PseudoDojo}. In addition, a Monkhorst-Pack grid in
momentum space with $10\times12\times1$ points was used, which yields well-converged
results in all cases. 

To obtain the 2D Fermi surfaces, an additional non self-consistent calculation was
performed for each stacking using the output results of the previous electronic
structure calculations. The only change in this step is the use of a denser M-K grid
with $50\times60\times1$ points. This finer calculation is then used to interpolate the bands of each stacking using Fourier interpolation approach, implemented in the BoltzTrap2 package \cite{BoltzTrap}.
An interpolation with 3 additional points for each interval was used in the previous
(non self-consistent) grid. The interpolated bands are finally post-processed to obtain
the contours at the Fermi level.




\subsection{Additional results}

To make the sliding process experimentally feasible, we must know if such configurations are stable. Fig.~\ref{fig:1} shows a comparison between stacking configurations and total energy variations in bilayer BPN systems. In panel (a) a pictorial atomic view of the stacking configurations is exhibited, beginning with the direct $(0.0, 0.0)$ stacking and sliding by $\delta_{y}$ and $\delta_{x}$ along the $y$ and $x$ axes, respectively. In panel (b),  
the energy variations (in meV/atom), $\Delta E=E_{\mu , \mu}-E_{0,0}$, are presented in a colored diagram for the different configurations. The  AA stacking $(0.0, 0.0)$ is found to have the smallest total energy, being the most stable configuration. However, the $\Delta E$ values of the other stackings are very close to each other ($\approx$ meV). This fact  suggests that the proposed configurations are feasible and stable, similar to bilayer graphene samples, for which the energy difference between several stackings is also of the order of meV \cite{Lee2016}. 

\begin{figure*}[t!]
    \centering
  \includegraphics[width=15cm, height=6cm]{(stackings+grafico_energias).jpg}
    \caption{(a) Atomic view of the stacking configurations starting from direct (0.0,0.0) and sliding $\delta_y$ and $\delta_x$, respectively, on top and bottom lines. (b) Energy variation $\Delta E$(meV/atom) diagram for the stacking configurations.}
    \label{fig:1}
\end{figure*}

\begin{figure*}[t!]
\includegraphics[width=2\columnwidth,trim={0 5cm 0 0},clip]{BPN_LSS_SG10_bands_QE (1).jpg}
\caption{Electronic bands for the LSS on line  $(\mu,0)$ (top row) and $(0,\mu)$ line (bottom row). In the case of the stacking $(0,0.4)$ the irreps indexation $\Delta_i$ at the bands are marked.}
\label{bands-LSS-10}
\end{figure*}

\begin{figure*}[t!]
\includegraphics[width=2\columnwidth,trim={0 5cm 0 0},clip]{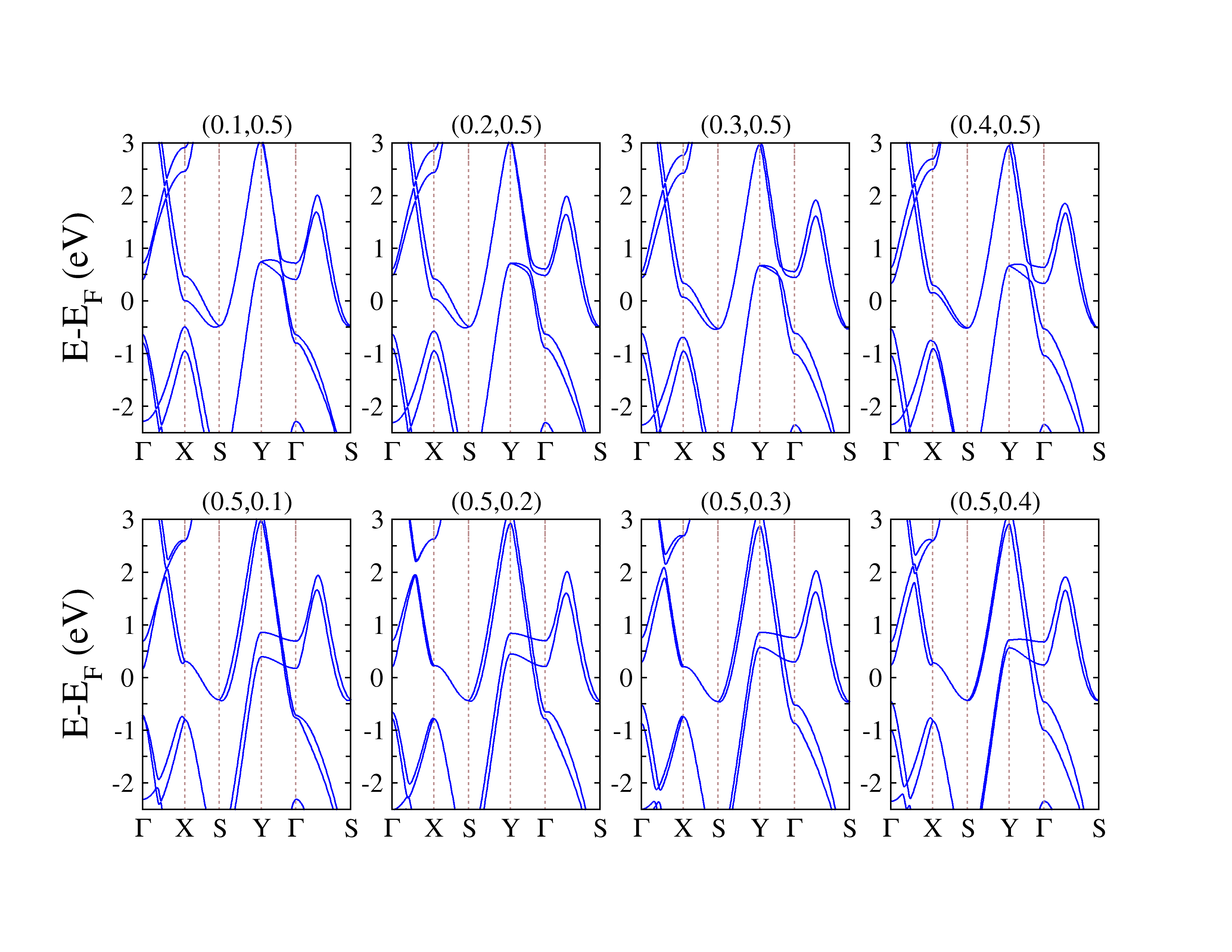}
\caption{Electronic bands for the LSS on lines $(\mu,0.5)$ and $(0.5,\mu)$ in the top and bottom rows, respectively.}
\label{bands-LSS-11}
\end{figure*}

\begin{figure*}[t!]
\includegraphics[width=2\columnwidth,trim={0 22cm 0 0},clip]{ 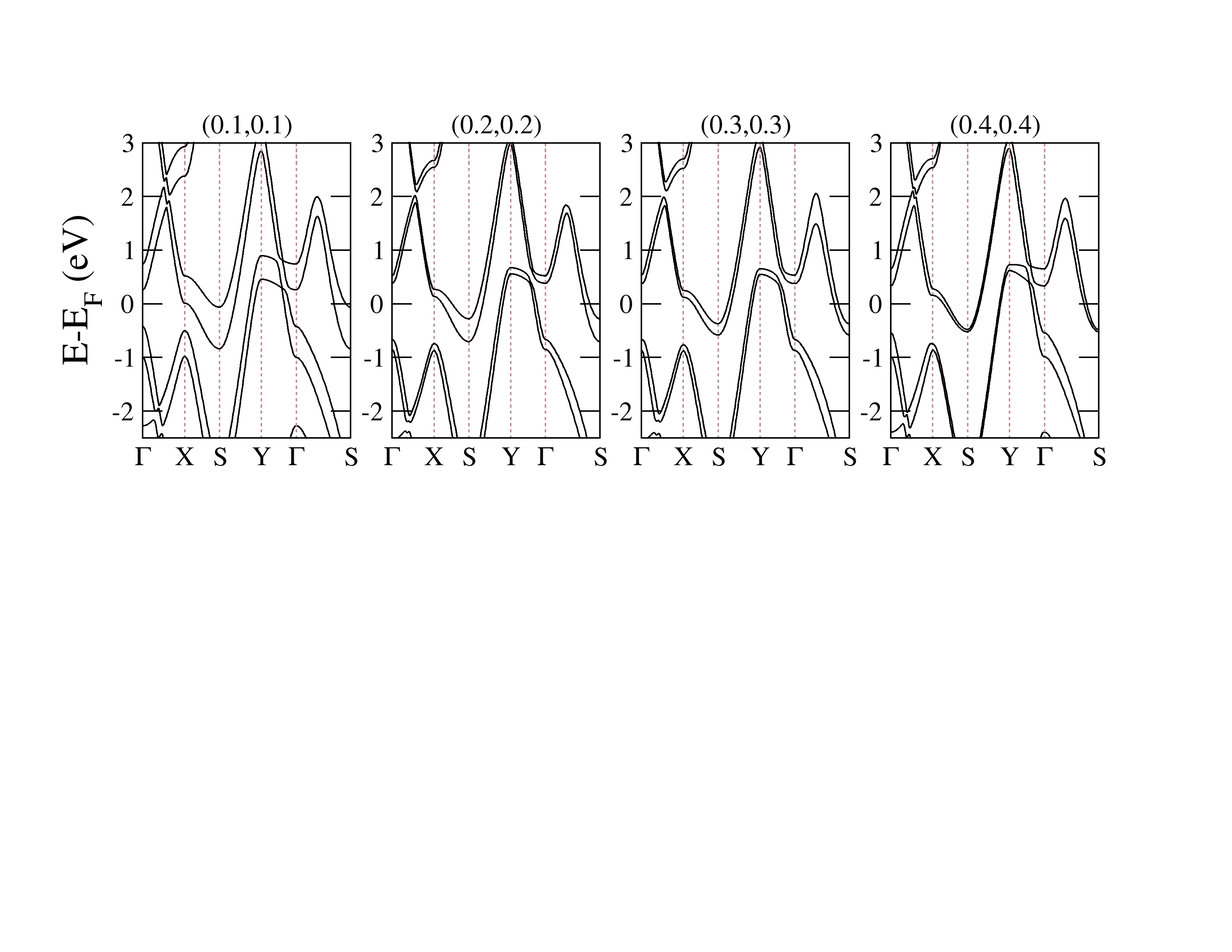}
\caption{Electronic bands for the LSS on line $(\mu,\mu)$.}
\label{bands-LSS-2}
\end{figure*}

A detailed picture of the band structure transitions across different sliding configurations is presented in Figs.~\ref{bands-LSS-10}--\ref{bands-LSS-2}. The colors represent the group symmetry adopted within the diagram presented in the main text. Three distinct scenarios are illustrated, corresponding to band configurations with different number of nodes.
The four-node configuration is observed along the $(0, \mu)$ line, 
shown in the second row of Fig.~\ref{bands-LSS-10}. In this case, both pairs of bands with similar electronic velocity share the same irreducible representations (irreps), as marked in the case of (0.,0.4) stacking. 
This irrep configuration remains consistent along the entire $(0, \mu)$ line, enabling a robust topological connection between the four Dirac cones and the high-symmetry points (HSS) located at $(0.0, 0.0)$ and $(0, 0.5)$. Along the $(\mu, 0)$ line, the band structure exhibits a qualitatively different evolution due to the specific grouping of irreps, changing from non-crossing band to two crossing  at approximately $\mu = 0.3$ [see first row of Fig.~\ref{bands-LSS-10}]. Other features are presented in Fig. S3, corresponding to the cases of $(\mu,0.5)$ and $(0.5,\mu)$ lines. Diagonal sliding geometries for the $(\mu,\mu)$ line are highlighted in Fig. S4.

\begin{figure*}
\includegraphics[width=2\columnwidth,clip]{ 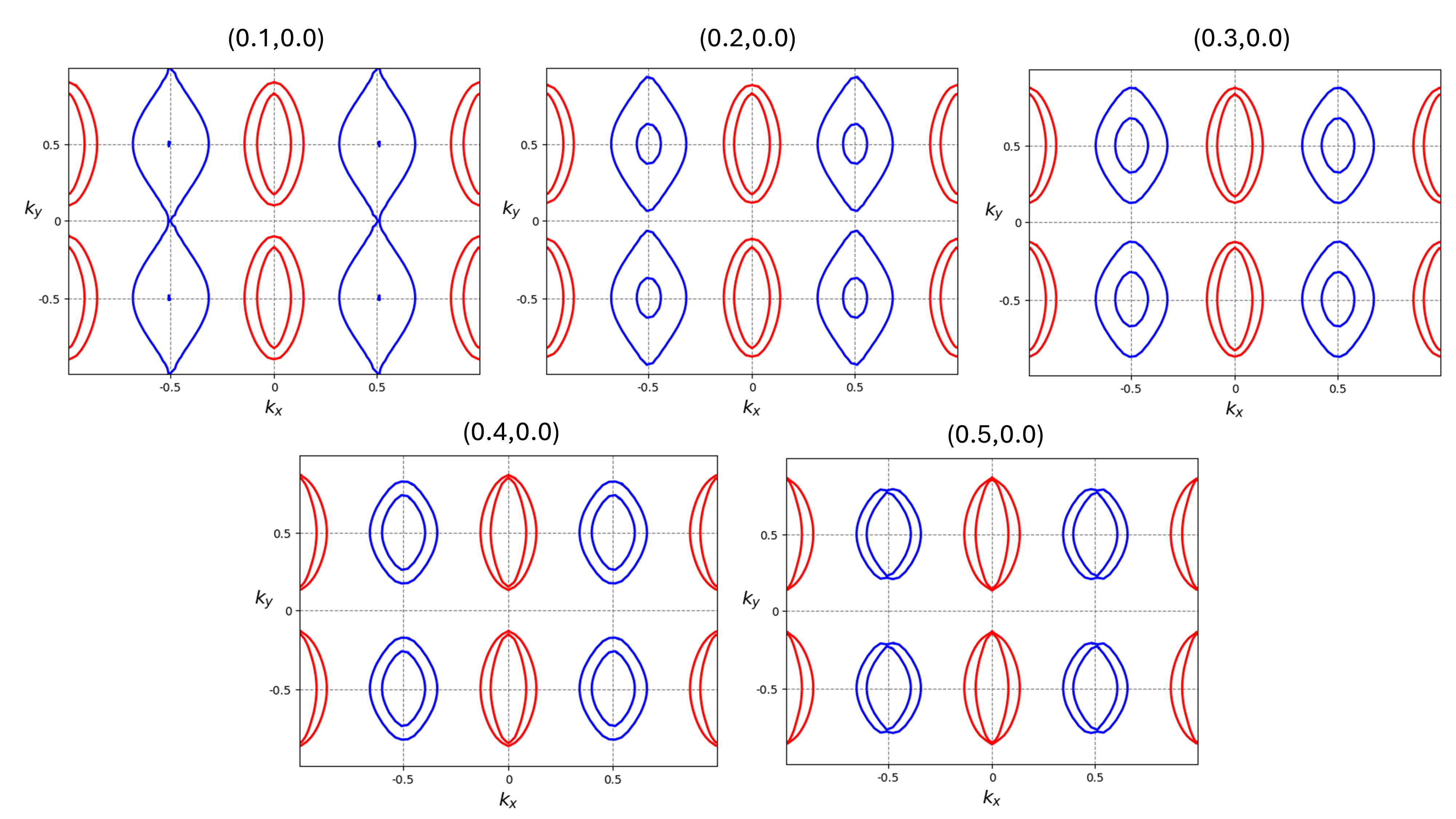}
\caption{Fermi surface (loops) for the $(\mu,0)$ line. Blue solid lines represent
electronlike surfaces while red solid lines correspond to holelike surfaces.}
\label{Fermi_surface_line_mu-0}
\end{figure*}

\begin{figure*}
\includegraphics[width=2\columnwidth,clip]{ 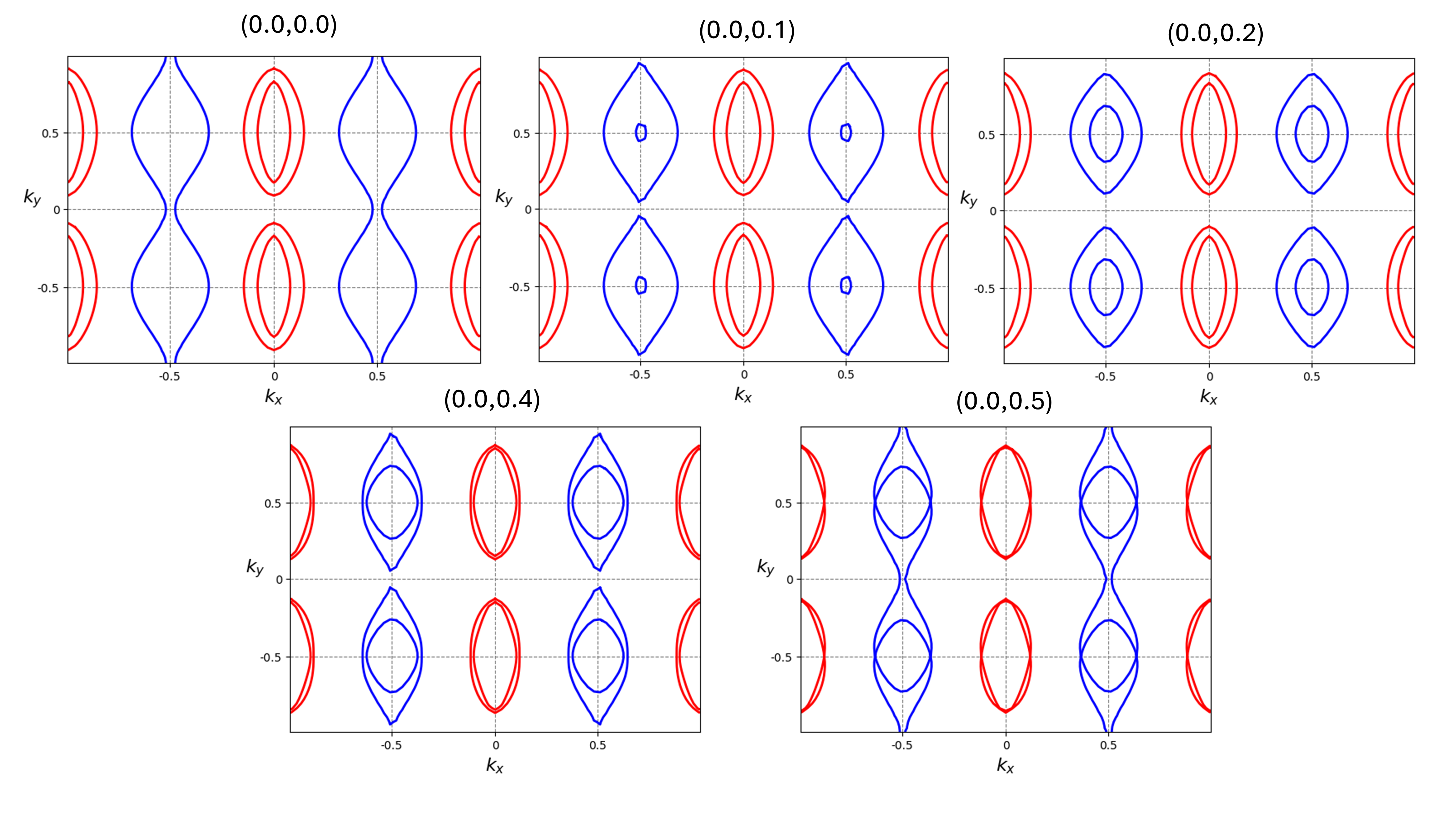}
\caption{Fermi surface (loops) for the $(0,\mu)$ line. Blue solid lines represent
electronlike surfaces while red solid lines correspond to holelike surfaces.}
\label{Fermi_surface_line_0-mu}
\end{figure*}

\begin{figure*}
\includegraphics[width=2\columnwidth,clip]{ 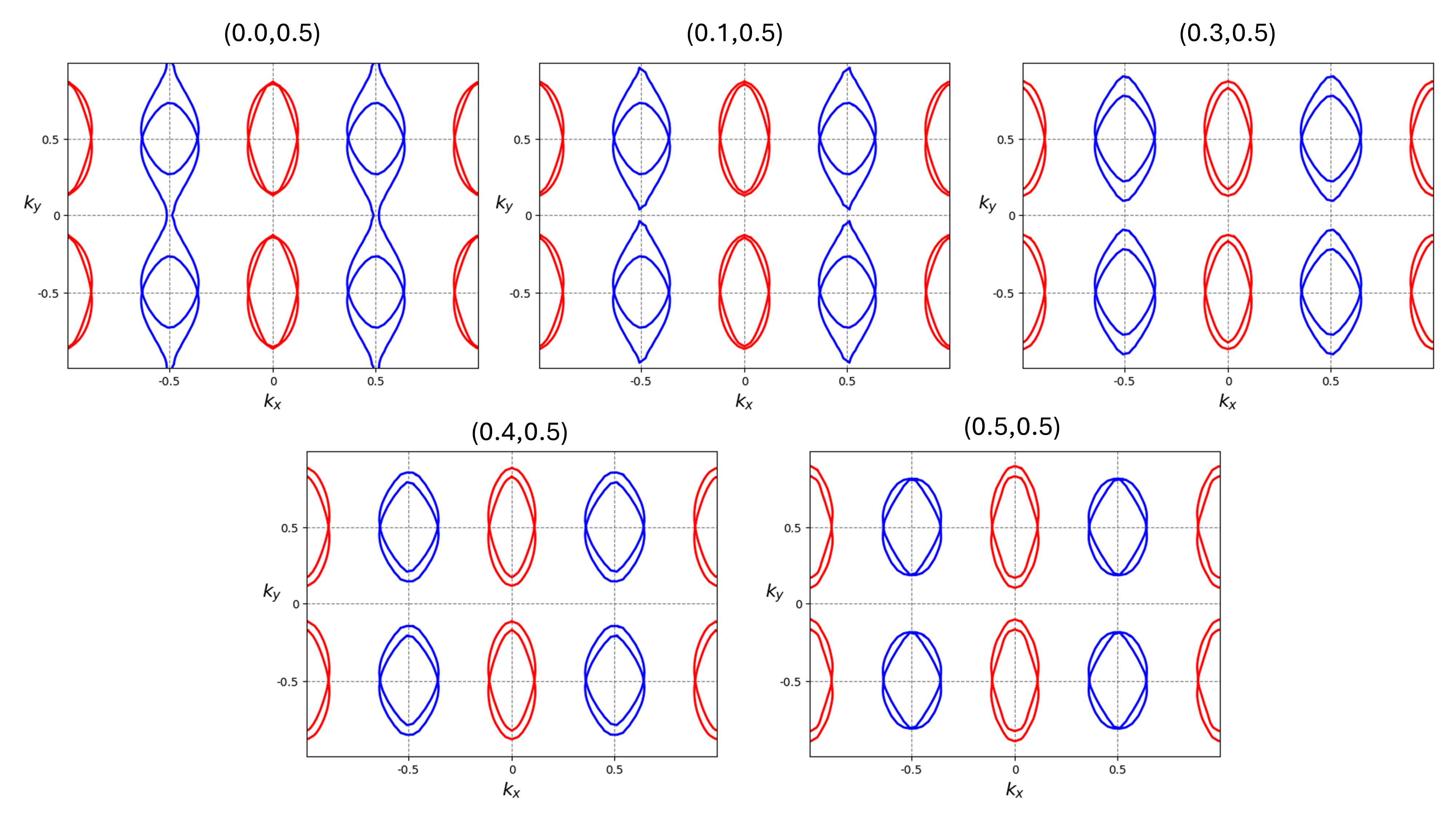}
\caption{Fermi surface (loops) for the $(\mu,0.5)$ line. Blue solid lines represent
electronlike surfaces while red solid lines correspond to holelike surfaces.}
\label{Fermi_surface_line_mu-0.5}
\end{figure*}

To conclude the discussion on the relationship between symmetry and electronic properties, Figs.~\ref{Fermi_surface_line_mu-0}--\ref{Fermi_surface_line_mu-0.5} present the contour plots for the displacements $(\mu, 0)$, $(0, \mu)$, and $(\mu, 0.5)$, respectively. In these plots, electronlike features are depicted in blue, while holelike features are represented in red. 

For the electronlike Fermi surfaces in Figs.~S6 and S7, which start from $(0.1, 0.0)$ and $(0.0, 0.0)$, respectively, a similar transition from open to closed loops is observed in both cases. However, near the final configuration at $\mu = 0.5$, distinct behaviors emerge. Specifically, at $(0.5, 0.0)$ in Fig.~S6, two closed loops are observed, while at $(0.0, 0.5)$ in Fig.~S7, the Fermi surface transitions back to an open loop, mirroring its initial configuration at $(0.0, 0.0)$. In contrast, the holelike Fermi surfaces exhibit opposite trends. For $(\mu, 0)$, both the initial and final configurations consist of two closed loops; however, for $(0, \mu)$, the final configuration at $\mu = 0.5$ results in a single closed loop. 

Finally, in Fig.~S8, the electronlike Fermi surface passes through a topological transition from an open loop to a single closed loop as it evolves from $(0.0, 0.5)$ to $(0.5, 0.5)$. In contrast, the holelike Fermi surface remains invariant within the sliding process. In summary, several cases exhibit topological transitions in the Fermi surfaces. For instance, the electronlike Fermi surfaces in Fig.~S6 undergo a transition from an open configuration to two closed loops at the point $(0.5, 0.0)$. Similarly, the holelike Fermi surfaces in Fig.~S7 transition from two closed loops to a single closed loop. Additionally, the electronlike Fermi surface in Fig.~S8 evolves from an open loop to a closed loop as it moves from $(0.0, 0.5)$ to $(0.5, 0.5)$.

In contrast, other cases remain invariant with the sliding process. The holelike Fermi surface in Fig.~S6 keep its configuration with two closed loops, showing no change. Similarly, the electronlike Fermi surface in Fig.~S7 still with the open-loop surface at  $(0.0, 0.5)$. Finally, the holelike Fermi surface in Fig.~S8 remains unchanged, consistently displaying a single closed loop throughout the changes in the stacking configurations.


\bibliography{refs}